\documentstyle[multirow,epsfig,amsmath,latexsym,amssymb]{mn}

\newcommand{\mi}{\text{min}}
\newcommand{\ma}{\text{max}}
\newcommand{\tB}{\text{B}}
\newcommand{\te}{\text{e}}
\newcommand{\td}{\text{d}}

\newlength{\MM} 
\newlength{\NN} 
\newlength{\OO} 
\newlength{\QQ} 
\newlength{\RR} 
\newlength{\TT} 

\begin{document}

\title[Visibility of GRB afterglows] 
{The visibility of gamma-ray burst afterglows in dusty star-forming regions} 

\author
[B.\,P. Venemans \& A.\,W. Blain] 
{
B.\,P.\ Venemans$^{1,2}$ and A.\,W.\ Blain$^3$\\
\vspace*{1mm}\\
$^1$ Leiden Observatory, P.O.\ Box 9513, 2300 RA Leiden, The Netherlands\\
$^2$ Cavendish Laboratory, Madingley Road, Cambridge CB3 0HE\\
$^3$ Institute of Astronomy, Madingley Road, Cambridge CB3 0HA\\
}
\maketitle

\begin{abstract}
Recent observations of the environments of gamma-ray bursts (GRBs)
favour massive stars as their progenitors, which are likely to be
surrounded by gas and dust. The visibility of the optical and UV
emission of a GRB are expected to depend on the characteristics of
both the dust and the GRB emission itself. A reasonable distribution
of surrounding dust is capable of absorbing all the optical and UV
emission of the optical flash and afterglow of a GRB, unless the
optical flash has a peak isotropic luminosity $L_{\text{peak}} \gtrsim
10^{49}\,$erg\,s$^{-1}$. This means that dark bursts should exist and
these bursts will have to be studied at infrared rather than optical
wavelengths. In this paper details will be given about the infrared
GRB dust emission. The reprocessed dust emission peaks at a rest-frame
wavelength of about $8\,\mu$m. Forthcoming space telescopes, in
particular the IRAC camera aboard the {\em Space Infrared Telescope
Facility}, could detect this emission out to a redshift of about
$2$. However, an accurate position of the GRB afterglow must be
provided for this emission to be identified, because the light curve
of the reprocessed dust emission does not vary on time-scales less
than several years.
\end{abstract}

\begin{keywords}
radiative transfer -- methods: observational -- ISM: clouds -- 
dust, extinction -- gamma-rays: burst -- infrared: galaxies
\end{keywords}

\section{Introduction}

The first gamma-ray bursts (GRBs), short and intense bursts of $0.1 -
1$\,MeV photons, were discovered at the end of the 1960s (Klebesadel,
Strong \& Olsen 1973). In the 1990s the BATSE detector on board the
{\em Compton Gamma-Ray Observatory}\footnote{see
http://cossc.gsfc.nasa.gov/cossc/cgro/batse\_src.html} observed
over 2500 GRBs and showed that GRBs were distributed isotropically over
the sky. A suspected conclusion was that GRBs were of extragalactic
origin at cosmological distances (see for example Paczynski 1995). This
conclusion was confirmed in 1997 using the {\em BeppoSAX} satellite by
the discovery of a well-located transient X-ray counterpart
(Costa et~al.\ 1997), and several hours later its association with a
fading optical point source (Groot et~al.\ 1997) at a redshift of
$0.695$ (Sahu et~al.\ 1997). Many more afterglows have now been
discovered and more redshifts have been measured (Greiner 2001). Based
on the observed gamma-ray flux densities, this implies that the total
energy radiated by a GRB is of order $10^{51}$ to $10^{53}$\,erg, if the
emission is isotropic, and is released in a few seconds.

The existence of multi-wavelength afterglows was predicted by models
that describe GRBs as opaque radiation-ionized fireballs, generating a
relativistic blast wave (e.g.\ M\'esz\'aros \& Rees 1993, 1997; Katz
1994); see Piran (1999) for an extensive review of the fireball
model. The most promising candidate objects that generate GRBs are
binary neutron star mergers (Piran 1999) and hypernovae or collapsars
(MacFadyen \& Woosley 1999; MacFadyen 2000 and references
therein). Observations of the environment in which GRBs occur
circumstantially favour massive stars as progenitors of GRBs. For
example, multiband photometric imaging of four GRB host galaxies show
their that spectral energy distributions (SEDs)  are best fitted by
starburst spectra (Sokolov et~al.\ 2001). Also, there are some
indications from {\em Hubble Space Telescope} observations that GRB
afterglows are associated with bright blue star-forming regions in
their host galaxies (for example Bloom et~al.\ 1999b; Fruchter
et~al.\ 1999; Bloom, Kulkarni \& Djorgovski 2001). The {\em Infrared Space Observatory} observed the
location of a single GRB, GRB970508, both weeks and months after the
burst and detected non-transient emission at $60\,\mu$m, which was
interpreted as an indication of the presence of an ultraluminous
infrared galaxy (ULIRG) (Hanlon et~al.\ 2000). If this ULIRG is the
host of the GRB, then the estimated star formation rate must be around
$200\,\rm{M}_{\odot}$\,yr$^{-1}$, likely placing this GRB in a very
powerful star forming region.  Another potentially important
discovery, which favours massive stars as progenitors of GRBs, is
evidence of a possible supernova component in the afterglow of
GRB970228 (Reichart 1999) and GRB980326 (Bloom et~al.\ 1999a). 
However, an alternative explanation for this, in which scattering of a
prompt optical flash by diffuse dust surrounding the GRB explains the
observations, has been proposed by Esin \& Blandford (2000).

If the progenitors of GRBs are indeed located in star-forming regions,
then the GRB is likely to be surrounded by gas and dust. This can influence
the spectrum of a GRB, and there have been several recent studies of
the effects. M{\'e}sz{\'a}ros \& Gruzinov (2000) showed that a dense
environment should modify the X-ray afterglow of a GRB, flattening the
X-ray light curve from hours to days after the burst, before
steepening the decay at later times. Waxman \& Draine (2000)
investigated dust  sublimation by GRBs and its implications. They
found that a prompt optical flash could destroy up to
$10^7$\,M$_{\odot}$ of dust out to $\sim 10\,$pc around a GRB. They
also calculated that the near-infrared luminosity from dust heated by the
optical flash could be as great as $10^{41}$\,erg\,s$^{-1}$.  Perna \&
Aguirre (2000) proposed to use GRB afterglows as probes of galactic
and intergalactic dust.  Recently, Galama \& Wijers (2001) found high
X-ray absorption column densities, but low optical extinctions for
their sample of GRB afterglows. They inferred that most GRBs are
embedded in large molecular clouds, and are therefore likely to be
produced by the death of short-lived massive stars, but that the
surrounding dust is destroyed by hard radiation from the GRB, leaving
the afterglow unextinguished.

In this article, the influence of surrounding dust on the visibility
of the afterglow and optical flash of a GRB is investigated in order
to study if such luminous events could be detected at high redshift,
even if they occur in dense dusty environments. The interaction of
homogeneously distributed dust around a GRB with the optical and UV
emission of the burst afterglow is modelled. The unextinguished light
curve of a GRB is described in section \ref{lcs}. A dust model
consisting of carbon grains with a range of sizes is used, which takes
account of the evolution of opacity, dust sublimation and temperature;
see section \ref{dustsec}. The model is used to estimate both the
visibility of the optical/UV afterglow for different densities and
profiles of the surrounding dust (section \ref{res1}), and the
properties of the thermal emission of this dust (section
\ref{res2}). Finally the flux densities of the dust emission are
compared with the sensitivity of future infrared detectors (section
\ref{obs}).

\section{GRB optical light curves}
\label{lcs}

\begin{figure}
\begin{center}
\epsfig{file=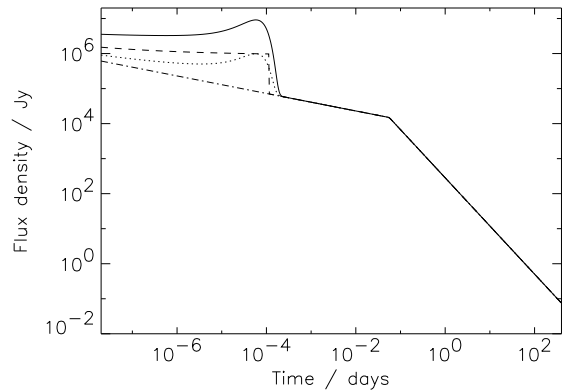}
\caption{\label{lc} B--band flux densities 1\,Mpc away from a
GRB with a light curve described by the model of Sari et~al.\ (1998)
used here. An afterglow with a Gaussian optical flash with
$L_{\text{peak}} = 10^{49}$\,erg\,s$^{-1}$ in the energy range $h\nu =
1 - 7.5$\,eV is shown by the solid curve; one with a Gaussian 
optical flash with $L_{\text{peak}} = 10^{48}$\,erg\,s$^{-1}$ is shown
by the dashed curve; an afterglow without an optical flash is shown by
the dot--dashed curve; and an afterglow with a top-hat optical flash with
$L_{\text{peak}} = 10^{48}$\,erg\,s$^{-1}$ is shown by the dotted curve.}
\end{center}
\end{figure}

The optical light curve of a typical GRB consists of two parts: a
prompt optical flash coincident with the gamma-ray emission and a
late-time afterglow. The afterglow can be explained by synchrotron
emission from a decelerating relativistic shell in collision with an
external medium: see Sari, Piran \& Narayan (1998) for an analytical
description of the light curve or Piran (1999) for a more detailed
derivation.

We assume that the peak frequency of the optical flash falls in
the optical waveband (Sari \& Piran 1999), and that the time profile
of the optical flash is represented as either a Gaussian or a top-hat
function. Both the peak luminosity and the duration of the flash are
left as free parameters, but typical values of the peak luminosity are
$L_{\nu} = 0.5 - 5 \times 10^{33}$\,erg\,s$^{-1}$\,Hz$^{-1}$ at a
frequency $\nu = 1.4 \times 10^{15}$\,Hz, with a SED $L_{\nu} \propto
\nu^{-1/2}$ and a duration of order 10\,s (Akerlof et~al.\ 1999;
Waxman \& Draine 2000). 

Waxman \& Draine (2000) showed that photons in the energy range $1 -
7.5$\,eV will dominate the heating of the dust, as in dense regions
(typically with hydrogen column densities $n_{\text{H}} \gtrsim
10^2$\,cm$^{-3}$) photons with energy $h\nu > 13.6\,$eV will
photoionize hydrogen very efficiently, while $7.5 - 13.6\,$eV photons
can be efficiently absorbed by H$_2$, by either vibrational excitation
or photodissociation (see also Draine 2000). The number of photons at
$h\nu > 7.5\,$eV that are available for dust heating will thus
probably be negligible. The lower limit of 1 eV is taken because both
the fraction of the energy radiated by the GRB below this energy is
small and the absorption efficiency of dust decreases rapidly at
wavelengths longer than about $1\,\mu$m ($\equiv 1.24\,$eV) (Draine \&
Lee 1984).

The following parameters are used to describe the afterglow, based on
values derived from observations (including Galama et~al.\ 1999a;
Lamb, Castander \& Reichart 1999; Huang, Dai \& Lu 2000; Greiner
2001). The energy of the shell $E = 5 \! \times \! 10^{52}\,$erg, the
fraction of the shell energy that goes into the kinetic energy of
electrons $\epsilon_{\te} = 0.15$, the ratio of the magnetic field
energy density to the total thermal energy $\epsilon_{\tB} = 0.1$, the
surrounding density $n_1= 5 \! \times \!  10^3\,$cm$^{-3}$ and the
coefficient of the power-law distribution of electron energies in the
GRB shell $p = 2.5$. The hydrodynamic evolution of the shock is
assumed to be adiabatic, because radiative evolution only applies if
$\epsilon_{\te} \simeq 1$.

In Fig.\ \ref{lc} four different optical light curves of GRBs, used
in our model, are plotted. 

\section{Radiation absorption and emission by dust grains}
\label{dustsec}

\subsection{Dust temperature}
\label{dusttemp}

The temperature of dust determines its emission spectrum. Because the
GRB afterglow is very luminous, the influence of heating by both
reprocessed dust emission from within the dust cloud and background
radiation from the interstellar radiation field of the galaxy can generally
be neglected. The dust grains are assumed to be spherical with a radius
$a$. A source of radiation with a luminosity $L = \int L_{\nu}
\td \nu$ is at a distance $r$ from a dust grain. If the temperature of a dust
grain is less than the temperature at which sublimation is a more
efficient energy loss mechanism than radiation, then the dust
temperature $T_{\td}$ can be calculated using

\begin{equation}
\label{eisewcv}
\pi a^2 \frac{L}{4 \pi r^2} \overline{Q_{\text{abs}}}(a) = 4 \pi
a^2~Q_{\text{em}}(a,T_{\text{d}}) \sigma T_{\text{d}}^4 +
\frac{4}{3} \pi a^3~C_{\text{V}} \frac{\td T_{\text{d}}}{\td t}  
\end{equation}

\noindent
(all the equations in this section are adapted from Draine \& Lee
1984, Siebenmorgen, Kr\"ugel \& Mathis 1992, Evans 1993 and Waxman \&
Draine 2000). $\overline{Q_{\text{abs}}}$ is the weighted effective
absorption efficiency, $\sigma$ is the Stefan-Boltzmann constant,
$C_{\text{V}}$ is the specific heat capacity and $Q_{\text{em}}$ is
the Planck-averaged emission efficiency,

\begin{equation}
\label{qemdef}
Q_{\text{em}}(a,T) \equiv \frac{\int B_{\nu}(T) Q_{\nu,\text{abs}}(a)
\td \nu}{\int B_{\nu}(T) \td \nu}.
\end{equation}

The weighted effective absorption efficiency

\begin{equation}
\overline{Q_{\text{abs}}} = \frac{\int_0^{\infty}~L_{\nu}
Q_{\nu,\text{abs}}~\td \nu}{\int_0^{\infty}~L_{\nu}~\td \nu}, 
\end{equation}

\noindent
depends on the spectrum of the source.

When subjected to very intense radiation from the optical flash of a
GRB, the temperature of dust can rise to levels where cooling of the
grains by sublimation is the dominant cooling process.

The cooling rate due to sublimation of a spherical dust grain per unit time,
$\dot{E}_{\text{sub}}$, 

\begin{equation}
\label{ecool}
\dot{E}_{\text{sub}} = 4 \pi a^2 \frac{\rho}{m} B \frac{\td a}{\td t}
\end{equation}

\noindent
can be described in terms of the reduction in the grain radius, $\td
a/\td t$, where $\rho$ is the mass density of the grain, $m$ the mean
atomic mass and $B$ is the chemical binding energy per atom within a
grain. An approximation to the sublimation rate, $\td a/\td t$, is given by

\begin{equation}
\label{grainsize}
\frac{\td a}{\td t} = - \frac{ P_0 e^{-T_0/T(t)}}{\rho} \left(\frac{m}{2
\pi k T} \right) ^{1/2}.
\end{equation}

$P_0$ and $T_0$ are constants, which depend on the properties of the
grain material. 

Now, using equation (\ref{eisewcv}) and including the
sublimation, equation (\ref{ecool}), the out-of-equilibrium temperature
$T_{\text{d}}$ of a dust grain can be calculated using

\begin{multline}
\label{graintemp}
\negthickspace \negthickspace
\negthickspace \pi a^2~\frac{e^{-\tau}~L}{4 \pi r^2}
\overline{Q_{\text{abs}}}(a) = \\ 
~4 \pi a^2~Q_{\text{em}}\left(a,T_{\text{d}}\right) \sigma
T_{\text{d}}^4 - 4 \pi a^2~\frac{\rho}{m} B \frac{\td a}{\td t} +
\frac{4}{3} \pi a^3~C_{\text{V}} \frac{\td T_{\text{d}}}{\td t},
\end{multline}

\noindent
in which $\tau$ is the optical depth between the source and the dust grain
(described in more detail in section \ref{tausec}). 
Equation (\ref{graintemp}) has the same structure as equation (8) of
Waxman \& Draine (2000), but with a term included for the heat capacity,
which can be important at temperatures below the sublimation
temperature, where thermal emission is the most important cooling mechanism.

Equations (\ref{grainsize}) and (\ref{graintemp})
are coupled differential equations for the evolution
of the temperature $T_d(t)$ and radius $a(t)$ of a dust grain.

The period of time during which a grain is entirely sublimated after
the arrival of a GRB afterglow is defined
as the destruction time and is, in general, a function of both grain
size and distance to the source: $t_{\text{des}} = t_{\text{des}}(a,r)$.

\subsection{Dust luminosity}
\label{dustemission}

The luminosity of a single dust grain is

\begin{equation}
L_{\text{d}}(a,t)=4 \pi a(t)^2 Q_{\text{em}}\left[a(t),T(t)\right]
\sigma T(t)^4 
\end{equation}

\noindent
assuming the grain radiates as a modified blackbody with an emission
efficiency $Q_{\text{em}}$ (equation \ref{qemdef}).

The flux density received a distance $r$ from a grain, that is
radiated in a frequency band between $\nu_{\mi}$ and 
$\nu_{\ma}$, can now be calculated. Defining
$x_{\mi}=h\nu_{\mi}/kT(t)$ and
$x_{\ma}=h\nu_{\ma}/kT(t)$, the flux at a distance $r$
from a single grain is approximately

\begin{equation}
\label{fluxa}
F(a,t) = \frac{1.9 \! \times \! 10^{-2}~L_{\text{d}}(a,t)}{4 \pi r^2} 
~\int_{x_{\mi}}^{x_{\ma}} \frac{{x'}^{4.5}}{e^{x'} - 1} \,\td x', 
\end{equation}

\noindent
assuming a reasonable emissivity index $\beta$ of $1.5$, where
$\epsilon_{\nu}~\propto~\nu^{\beta}$.  To calculate the evolution of
the flux of a shell with radius $R$ consisting of grains with size
$a$, the flux of all the grains must be added, taking into account the
time delay due to the differences in distance from the individual
grains in the shell to the observer. A photon emitted by a grain at an
angle $\theta$ from the line of sight has to travel an extra distance
of $R\,(1-\cos\theta)$ as compared with a photon emitted from the same
shell at the same time, but directly along the line of sight in order
to be detected simultaneously by the same observer. If we assume that
the distance to the observer is much greater than the radius of the
shell ($r \gg R$), and the number density of grains is $n$, then the
total flux received by the observer is the integral over all shells:

\begin{multline}
\label{ftot}
\negthickspace \negthickspace \negthickspace
\negthickspace F_{\text{tot}}(t) = \int_a \int_R F(R,a,t) \,\td R \,\td a = \\
\int_a \int_R \int_{\theta} 2 \pi R^2 \,n(R,a) \sin \theta \,
F[a,t+\frac{R}{c}(\cos \theta -1)] \, \td \theta \,\td R \,\td a.
\end{multline} 

Equation (\ref{ftot}) can only be used for calculating the dust emission
in the case of a spherical distribution of dust around a GRB, the
emission from which is itself isotropic. If, for example, the
afterglow emission of the GRB is beamed, then (roughly) only the dust
inside the opening angle $\theta_0$ of the beam will be heated. In
that case, the emission will be reduced by $\sim\theta_0^2/4$, and
the time profile of the dust radiation will change, depending on both the
opening angle and inclination of the beam to the line of sight. Recently,
Frail et al.\ (2001) derived opening angles of GRBs by analysing
gradient changes in the multicolour light curves of GRB
afterglows. They concluded that the afterglow emission of GRBs is
generally beamed, with values for $\theta_0$ ranging between $0.05$
and $0.4$ radians. This could hence be an important effect, significantly
reducing the intensity of the infrared radiation emitted from a
reprocessed afterglow as compared with the predictions in the
isotropic model presented here. 

In a preprint pointed out by the referee, Fruchter, Krolik \& Rhoads
(2001) discuss dust destruction by X-rays. They found that heating by
the optical/UV flash and by X-rays have roughly comparable importance
for grain evaporation, and either may be more important, depending on
the burst spectrum and grain size distribution. Optical/UV heating is
generally more important for small grains and X-ray heating for large
grains, although the division depends on grain composition. They also
found that under most circumstances grain charging and electrostatic
shattering is a more effective mechanism in destroying dust grains,
compared to heating and evaporation, in disagreement with the
conclusions of Waxman \& Draine (2000), who found that grain
shattering does not appreciably reduce the UV/optical extinction. The
processes described by Fruchter et~al.\ (2001) could reduce the number
of dust grains surrounding a GRB, and so further reduce the infrared
dust emission.

If the distribution of the dust is not homogeneous but clumpy, then the
heating of a dust grain will not only depend on $R$ but also on
$\theta$ and $\phi$ ($F = F[R,a,t,\theta,\phi]$) and equation
\ref{ftot} will no longer be valid. At some places, radiation levels
inside the clumpy medium are greatly enhanced with respect to a
homogeneous medium due to relatively dust-free lines of sight, while
at other places radiation levels are lower due to more effective
screening (for example, see Hobson \& Padman 1993; Hobson \& Scheuer 1993). 

\subsection{Optical depth}
\label{tausec}

Assume a source of radiation is surrounded by spherical
shells of dust. The optical/UV flux from the source incident on a
dust grain at a given radius from the source will be attenuated by the
dust at smaller radii, and so the optical depth, $\tau$, 
between the source and the grain will be non zero. In general, $\tau_{\nu}$
is defined as 

\begin{equation}
\label{tau1}
\tau_{\nu} = \int_r \kappa_{\nu} ~\td r = \int_r \int_a n(a,r) \sigma_{\nu}(a)
~\td a\, \td r
\end{equation}

\noindent 
where $\sigma_{\nu}(a)$ is the absorption cross-section of a dust grain with
size $a$ and $n(a,r)$ is the number density of dust grains with sizes
between $a$ and $a+\text{d}a$ at a distance $r$ from the
GRB. Because some of the dust grains can be sublimated due to intense
heating from the source, $\tau_{\nu}$ will in general also be a function
of time. The absorption cross section is

\begin{equation}
\label{tauart}
\sigma_{\nu}(a,r,t) = 
\begin{cases}
Q_{\nu,\text{abs}}(a) \pi a^2, & t < t_{\text{des}}(a,r) \\
0, & t > t_{\text{des}}(a,r), \\
\end{cases}
\end{equation}

\noindent
where $t_{\text{des}}(a,r)$ is the destruction time of a grain with
size $a$ at a distance $r$ from the source and $Q_{\nu,\text{abs}}(a)$
the absorption coefficient. To compute $\kappa_{\nu}$ an expression for
$n(a,r)$ is required. In this model the empirical law

\begin{equation}
\label{nar}
n(a,r) = C(r) a^{-3.5}
\end{equation} 

\noindent
(Mathis, Rumpl \& Nordsieck 1977; Hildebrand 1983) was used for the
grain size distribution, in which

\begin{equation} 
\label{cr} 
C(r) =
\frac{3~n_{\text{H}}(r)~m_{\text{H}}}{8\pi\rho\,
M_{\text{g}}/M_{\text{d}}\,\left(\sqrt{a_{\text{max}}} -
\sqrt{a_{\text{min}}}\right)}.  
\end{equation}

\noindent
$n_{\text{H}}(r)$ is the number density of neutral hydrogen,
$m_{\text{H}}$ is the atomic mass of hydrogen,
$M_{\text{g}}/M_{\text{d}}$ is the gas-to-dust mass ratio,
$\rho$ is the mean mass density of the dust grains, and $a_{\text{min}}$ and
$a_{\text{max}}$ are the minimum and maximum grain radii. 

Using equations (\ref{tau1}) to (\ref{cr}), the total optical depth
can now be calculated. This is the optical depth due to dust
immediately surrounding the GRB. In addition, an afterglow light curve
can be attenuated by dust along the line of sight within the host
galaxy of the burst, in the intergalactic medium (Perna \& Aguirre 2000)
and in our own galaxy. 

With the expressions for the light curve of GRB afterglows with or
without an optical flash (see section \ref{lcs} above), the luminosity in
the photon energy range $h\nu = 1 - 7.5\,$eV, $L_{1-7.5}(t)$, can now
be calculated, and so the dust properties $T(t)$, $\tau(r,t)$ and
$a(t)$ can be computed throughout the enshrouding dust cloud, as the
optical emission of the GRB travels through it.

\subsection{Numerical model}
\label{nummod}

\subsubsection{Grain material}

For the simulations a model of dust  consisting of carbon grains with
various sizes was used. Graphite grains are better absorbers of
optical and UV photons than silicate grains (Mezger, Mathis \& Panagia 1982;
Draine \& Lee 1984). On the other hand, silicate grains are destroyed
at lower temperatures than graphite grains because their chemical bonds are
weaker (Guhathakurta \& Draine 1989; Evans 1993). Graphite grains will
thus get hotter, but are also more robust, and so the destruction time
of graphites and silicates are not expected to differ by large
factors. Hence, because of their higher absorptivity, graphite grains
are likely to dominate the optical depth at optical and UV wavelengths.

\subsubsection{Absorption and emission coefficients}

The value of the absorption coefficient of a graphite grain in the
energy range $1 - 7.5\,$eV, $\overline{Q_{\text{abs}}}$, depends on the
grain size. Based on the results of computations by Draine \& Lee (1984),
if we define an effective wavelength $\lambda_{\text{eff}}$ to account
for the interaction of the physical properties of the grains and the
illuminating radiation field, then the absorption coefficient 

\begin{equation}
\overline{Q_{\text{abs}}}(a) = 
\begin{cases}
10\, a\, \lambda_{\text{eff}}^{-1}, & a < 0.1\, \lambda_{\text{eff}} \\
1, & a > 0.1\, \lambda_{\text{eff}}.
\end{cases}
\end{equation}

A value of $0.5\,\mu$m is taken for
$\lambda_{\text{eff}}$, because in the GRB afterglow model
approximately half the energy is radiated at longer wavelengths and
half at shorter wavelengths, depending on whether or not there is an
optical flash, which is expected to be bluer than a typical
afterglow. Changing $\lambda_{\text{eff}}$ from $0.4\,\mu$m to
$0.6\,\mu$m made little 
difference to the outcome of the simulations: the destruction time of
the grains and the dust emission both changed by only about 1
per cent. Empirical values from Draine (2001) were taken for the
emission coefficients, $Q_{\text{em}}$ (equation \ref{qemdef}). 

\subsubsection{Density profiles and dust mass}

Two different dust density profiles were considered; a constant
density at all radii and an isothermal density profile, that is $n(r)
\propto r^{-2}$. A gas-to-dust mass ratio of $10^2$ was assumed
(Hildebrand 1983; Evans 1993). 
In the model it is assumed there is no dust within 0.3 pc of the
GRB. This is the Str\"{o}mgren radius of an HII region with
$n_{\text{H}} = 5000$\,cm$^{-3}$ surrounding a star of spectral type
O6. If the progenitors of GRBs are massive stars of this type, then it
is unlikely that significant amounts of dust will be able to survive
within the Str\"{o}mgren sphere. Assuming a different inner radius, however,
does not affect the optical depth or the dust mass strongly. Even if
the inner radius is reduced to zero, then the dust mass will 
change by only a few percent. The maximum radius is taken
to be 3.67\,pc and 10.33\,pc in the case of a constant and isothermal
density profile respectively. The total dust mass is 
0.234\,($n_{\text{H}_{0.3}}/$cm$^{-3}$)\,M$_{\odot}$ and
0.016\,($n_{\text{H}_{0.3}}/$cm$^{-3}$)\,M$_{\odot}$ in these cases
respectively, where the density of hydrogen at 0.3
pc ($n_{\text{H}_{0.3}}$) is a free parameter in the model.

\subsubsection{Minimum and maximum grain size}

As mentioned in section \ref{tausec} an empirical law (equation
\ref{nar}) was used for the distribution of dust grain sizes from
$0.005 \,\mu$m ($a_{\mi}$) to $2.5 \,\mu$m ($a_{\ma}$). The minimum
grain size was derived by Mathis et~al.\ (1977). The maximum grain
size is adopted from Witt, Oliveri \& Schild (1990) and Aitken et al.\
(1993). However, the maximum grain size assumed by different authors
varies widely, in papers about theory as well as in papers about
observations.  To investigate the effect of the minimum and maximum
value of the grain size on the results of the dust model, the range
was extended to $0.001\, \mu$m as a lower limit, to reach molecular
sizes, and to $5 \,\mu$m as an upper limit.

\begin{table}
\caption{\label{aandtau} Optical depth all the way through the dust cloud
for different profiles and limits of grain sizes. The total dust mass
and maximum radius in the case of a flat density profile is
$0.234\,(n_{\text{H}_{0.3}}$/cm$^{-3})$\,M$_{\odot}$ and 3.67\,pc. In
the case of decreasing profile these are
$0.016\,(n_{\text{H}_{0.3}}$/cm$^{-3})$\,M$_{\odot}$ and 10.33\,pc.
To calculate the optical depth at other wavelengths,
it is assumed that the optical depth varies as $\propto
\lambda^{-1}$ at wavelengths $0.5\,\mu$m $< \lambda < 1.25\,\mu$m (for
example, Evans 1993) and as $\propto \lambda^{-1.6}$ at wavelengths $\lambda
> 1.25\,\mu$m (Cardilli, Clayton \& Mathis 1989; Mathis 1990).}
\vspace{0.3cm} 
\begin{tabular}{|c|c|c|c|}
\hline
profile & $a_{\mi}$/cm & $a_{\ma}$/cm  
& $\tau(\lambda = 0.5\,\mu$m$)$/cm$^{-3}$ \\
\hline
\multirow{2}{\QQ}{flat} & {$5 \! \times 10^{-7}$} & {$2.5 \! \times
10^{-4}$} & {$2.8 \! \times \! 10^{-3} ~n_{\text{H}_{0.3}}$} \\   
\multirow{2}{\QQ}{$\propto r^0$} & $10^{-7}$ & $2.5 \! \times
10^{-4}$ & {$3.1 \! \times \! 10^{-3}~ n_{\text{H}_{0.3}}$} \\  
{} & $5 \! \times 10^{-7}$ & $5 \! \times 10^{-4}$ & {$1.8 \! \times
\! 10^{-3}~n_{\text{H}_{0.3}}$} \\ 
\noalign{\medskip}
\multirow{2}{\QQ}{isothermal} & $5 \! \times 10^{-7}$ & $2.5 \! \times
10^{-4}$ & {$2.6 \! \times \! 10^{-4} ~n_{\text{H}_{0.3}}$} \\  
\multirow{2}{\QQ}{$\propto r^{-2}$} & $10^{-7}$ & $2.5 \! \times
10^{-4}$ & {$2.9 \! \times \! 10^{-4} ~n_{\text{H}_{0.3}}$} \\ 
{} & $5 \! \times 10^{-7}$ & $5 \! \times 10^{-4}$ & {$1.7 \! \times \!
10^{-4} ~n_{\text{H}_{0.3}}$} \\ 
\hline
\end{tabular}
\end{table}

The upper limit to the grain size has the strongest influence on the
total number of dust grains for a fixed dust mass in the model,
because the maximum grain size determines the value of $C(r)$
(equation~\ref{cr}). If the maximum size of the grains is increased by
a factor of 2, then the value of $C(r)$ will decrease by a factor of
approximately $\sqrt{2}$. This influences both the optical depth and
the dust luminosity; for example, for the same total
dust mass, the decrease in the number of small grains as $a_{\ma}$
increases will lead to a reduction in the optical depth, because the
optical depth is dominated by smaller grains that are still larger
than molecules
(see equation \ref{tauart}: d$\tau \sim a^{-1.5}$ d$a$), as shown in
Table\,\ref{aandtau}. Hence, the choice of  maximum grain size has a
significant effect, because it determines the number of
dust grains. However, if $C(r)$, and thus the number of grains with a
given radius, is kept fixed, then neither the optical depth nor the
dust emission change significantly as a function of $a_{\ma}$: the change in
dust emission with $a_{\ma}$ is less than $1$ per cent at all
wavelengths. The effect of changing the upper limit of grain
size in the model is thus negligible if the normalization of $n(a,r)$ is
kept constant, although the total mass of dust will then change.

Decreasing the lower limit of the grain size from $0.005\,\mu$m to
$0.001\,\mu$m with $n_{\text{H}_{0.3}}$ fixed increases the optical
depth by about $11$ per cent (see Table\,\ref{aandtau}), while $C(r)$
only changes by about $2$ per cent. Because of the increased amount
of energy absorbed by the smallest grains when $a_{\mi}$ is reduced,
the optical/UV flux density of the late-time afterglow is also reduced
by 40 per cent.
However, the dust luminosity does not change by a
large amount, because the fractional change in the amount of energy
absorbed is only of order 0.1 per cent. Decreasing the minimum grain
size further has a negligible effect on both the dust
emission and the optical depth, as the additional amount of energy
absorbed is very small.

Absorption by large molecules or Polycyclic Aromatic Hydrocarbon
maolecules (PAHs) only become an important
factor for the optical depth at photon energies $h\nu \gtrsim 10$\,eV,
at which their absorption cross sections increase to be comparable to
their physical cross section
(Draine \& Lee 1984). Below those photon energies, the optical
depth is dominated by bigger grains (see also D\'esert, Boulanger \& Puget
1990). Because photons with energy $h\nu > 7.5$\,eV will mainly be
absorbed both by H and H$_2$ in the early phases of the GRB (Waxman \&
Draine 2000, see section \ref{lcs}), the smallest grains and molecules
are unlikely to contribute significantly to the total dust emission.

The minimum and maximum grain sizes used in the model were
assumed to be $0.005\,\mu$m and $2.5\,\mu$m respectively. However,
changing the minimum and maximum grain size was found not 
to influence the results of the model significantly, if the
number of grains with a given radius was kept fixed.

\begin{table}
\caption{\label{sims} Values of some important parameters, used for
simulations. In the table OF is the representation of the optical
flash (e.g. Gaussian), {$L_{\text{peak}}$} is the peak luminosity
of the optical flash in the energy band $1 - 7.5$\,eV in erg\,s$^{-1}$ 
and profile is the density profile. The duration of the optical flash
is in each case 10\,s. In Fig.\,\ref{lc} the unattenuated light curves
are plotted.} 
\begin{tabular}{|c||c|c|c|c|}
\hline
{light curve} & {$L_{\text{peak}}$} & 
{$n_{\text{H}}$/cm$^{-3}$} & {profile} & {name} \\ 
\hline
{} & {} & {$5 \! \times \! 10^3$} & {flat} & {$A_1$} \\ 
\multirow{2}{\RR}{Gaussian OF +} & {} & {$5 \! \times \! 10^4$} &
{flat} & {$A_2$} \\ 
\multirow{2}{\TT}{afterglow} & {$10^{49}$} & {$5 \! \times \! 10^3$} &
{isothermal} & {$A_3$} \\ 
{} & {} & {$5 \! \times \! 10^4$} & {isothermal} & {$A_4$} \\ 
{} & {} & {$5 \! \times \! 10^5$} & {isothermal} & {$A_5$} \\ 
\noalign{\medskip}
{} & {} & {$5 \! \times \! 10^3$} & {flat} & {$B_1$} \\ 
\multirow{2}{\RR}{Gaussian OF +} & {} & {$5 \! \times \! 10^4$} &
{flat} & {$B_2$} \\ 
\multirow{2}{\TT}{afterglow} & {$10^{48}$} & {$5 \! \times \! 10^3$} &
{isothermal} & {$B_3$} \\ 
{} & {} & {$5 \! \times \! 10^4$} & {isothermal} & {$B_4$} \\
{} & {} & {$5 \! \times \! 10^5$} & {isothermal} & {$B_5$} \\ 
\noalign{\medskip}
{} & {} & {$5 \! \times \! 10^3$} & {flat} & {$C_1$} \\ 
{} & {} & {$5 \! \times \! 10^4$} & {flat} & {$C_2$} \\ 
{afterglow only} & {--} & {$5 \! \times \! 10^3$} & {isothermal} & {$C_3$} \\ 
{} & {} & {$5 \! \times \! 10^4$} & {isothermal} & {$C_4$} \\ 
{} & {} & {$5 \! \times \! 10^5$} & {isothermal} & {$C_5$} \\ 
\noalign{\medskip}
{Top hat OF +} & \multirow{2}{\NN}{$10^{48}$} &
\multirow{2}{\MM}{$5 \! \times \! 10^5$} &
\multirow{2}{\QQ}{isothermal} & \multirow{2}{\OO}{$D$} \\ 
{afterglow} & {} & {} & {} & {} \\
\hline
\end{tabular}
\end{table}

\section{Results: optical light curves}
\label{res1}

\begin{figure}
\epsfig{file=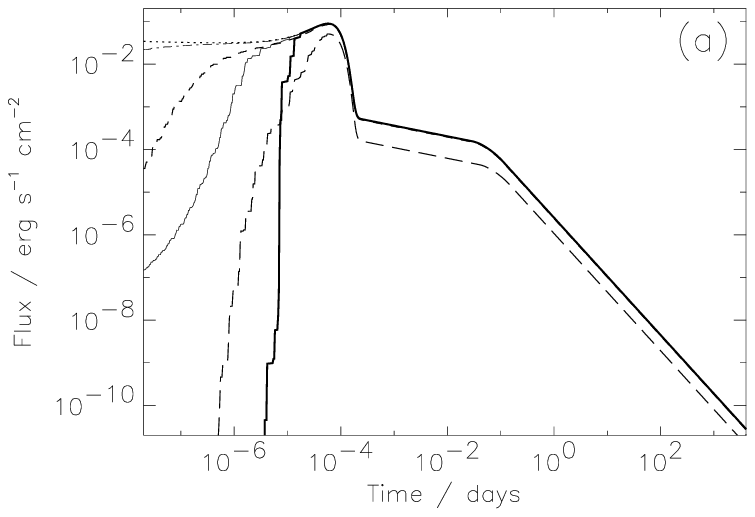}
%\vspace{-0.5cm}
\epsfig{file=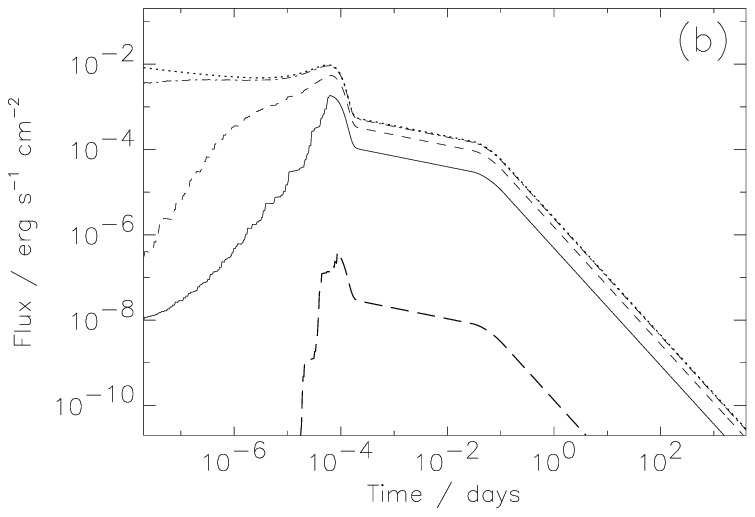}
%\vspace{-0.5cm}
\epsfig{file=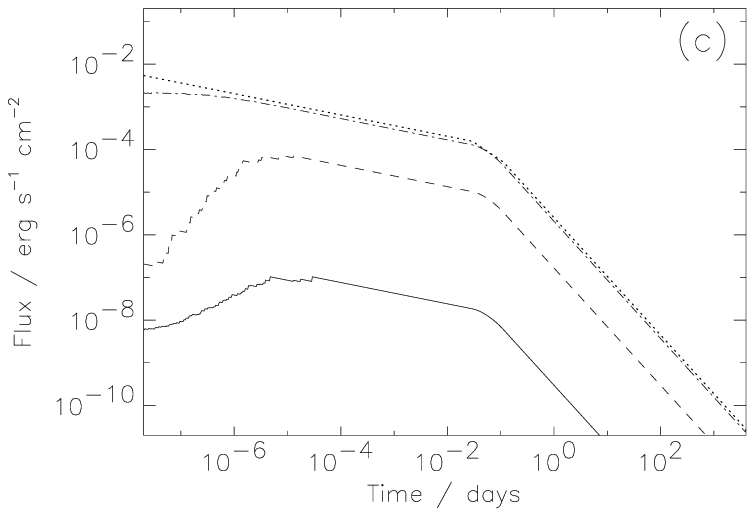}
\caption{\label{alllc} (a) Integrated flux of a GRB afterglow plus
a Gaussian optical flash with L$_{\text{peak}} = 10^{49}\,$erg$\,$s$^{-1}$
at a distance of 1 Mpc between $\nu = 2.4 \times 10^{14}$\,Hz and $\nu
= 1.8 \times 10^{15}$\,Hz. The dotted line is the light curve with no
extinction. The thin solid line and the thick solid line include absorption
by dust with a flat density profile and a density of $n_{\text{H}} = 0.5$
and $5 \! \times \! 10^4\,$cm$^{-3}$ respectively (light curve $A_1$
and $A_2$ of Table\,\ref{sims}). The dash-dotted
line, the dashed line and the long-dashed line include
absorption by dust with an isothermal density profile and a maximum
density of $n_{\text{H,max}} = 0.5$, $5$ and $50 \! \times \! 10^4\,$cm$^{-3}$
respectively (light curve $A_3$, $A_4$ and $A_5$ of
Table\,\ref{sims}). (b) As (a), but with an optical flash
with L$_{\text{peak}} = 10^{48}\,$erg$\,$s$^{-1}$. The
light curve which includes absorption by dust with constant density of
$n_{\text{H}} = 5\!\times\!10^4\,$cm$^{-3}$ (light curve $B_2$) is
not shown, because the flux is close to zero because of the very powerful
absorption in this case. (c) As (a), but with no optical flash. 
Light curves $C_2$ and $C_5$ are not shown, 
because the flux is almost zero due to absorption by the dust.}
\end{figure}

In Table\,\ref{sims} the parameters describing the light curves and
the properties of surrounding dust clouds used here are listed.
In Fig.\,\ref{alllc}a light curves are plotted for a GRB with an
afterglow and an optical flash with a peak luminosity of
$10^{49}$\,erg\,s$^{-1}$ as it would appear to an observer at a
distance of 1\,Mpc, including the effects of absorption by dust
surrounding the GRB. 

As can be seen in Fig.\,\ref{alllc}a, the optical flash is able to
destroy large amounts of dust. Although the total dust mass in the
cloud varies a great deal from model to model (see section \ref{nummod}),
and can be as great as $2300$\,M$_{\odot}$, the visibility of the
afterglow is hardly reduced after the first few seconds. In the case
of a flat dust density profile with $n_{\text{H}} = 0.5$ and $5 \! \times \!
10^4$\,cm$^{-3}$, as represented by curve $A_1$ and $A_2$ respectively, all 
the dust is destroyed by the optical flash and observations of the
afterglow would not reveal any dust surrounding the
GRB. In the case of an isothermal density profile, some of the bigger
dust grains in the outer shells survive the passage of the flash. This is
not because of screening by the absorption due to grains at smaller
radii, but because the distance to the burst is sufficiently large
that the intensity of the optical/UV flux has decreased below the
level at which grains are destroyed by sublimation. For example,
beyond about 4.5\,pc the flux density of even an unattenuated flash with
a peak luminosity of $10^{49}$\,erg\,s$^{-1}$ is not sufficient to
sublime the largest grains ($a \gtrsim 1\,\mu$m).

When the peak of the optical flash is less intense,
$10^{48}$\,erg\,s$^{-1}$, less dust is destroyed along the line of
sight, and so the afterglow is attenuated to a greater degree
(see Fig.\,\ref{alllc}b). In the case of a flat profile with
$n_{\text{H}} = 5\!\times\!10^4$\,cm$^{-3}$, practically all the
optical/UV energy of the flash and afterglow is absorbed by the
surrounding dust. In the case of an isothermal profile and a density
of $n_{\text{H}} = 5 \! \times \! 10^5$\,cm$^{-3}$, the afterglow is
attenuated by a factor of about $2.5 \! \times \! 10^5$ or 11
magnitudes. Hence, it is very unlikely that an optical/UV afterglow would be
detected; for example, the unattenuated R-band magnitude of a GRB
afterglow at a relatively low redshift $z = 0.3$ one day after the
start of the burst, which is predicted to be $R \sim 18$, would be reduced
to $R \sim 29$.

In the densest environments, the afterglow of a GRB with a relatively
weak optical flash will be heavily attenuated; however, in less dense
environments, such a burst will only be marginally affected.

If there is no optical flash (or if the intensity of the optical
flash has an insignificant value of $L_{\text{peak}} \lesssim
10^{47}$\,erg\,s$^{-1}$), then almost no dust is destroyed, and so the
optical/UV afterglow emission from the GRB will be severely attenuated
in all cases: see Fig.\,\ref{alllc}c. If the shape of the optical
flash is a top-hat and not a Gaussian, then more dust will be
destroyed, simply because there is more energy in a flash with a
top-hat profile than in a Gaussian flash with the same peak flux.

In conclusion, the higher the peak luminosity of the optical flash,
the greater is the amount of dust that is expected to be destroyed in
the environment of the GRB, and so the afterglow will be less
attenuated. If the peak luminosity of the optical flash is $\gtrsim
10^{49}$\,erg\,s$^{-1}$, then the intensity of the afterglow emission
will never be reduced significantly.  If the peak luminosity of the
optical flash $\lesssim 10^{47}$\,erg\,s$^{-1}$, then the afterglow
will always be highly obscured, unless the total mass of the dust
within a few pc surrounding the GRB is less than $100$\,M$_{\odot}$,
which would be the case for $n_{\text{H}} \lesssim 10^3\,$cm$^{-3}$
and a gas-to-dust mass ratio of about $10^2$.

\section{Results: dust emission}
\label{res2}

\subsection{Dust emission from the GRB environment} 
\label{onesource}

\begin{figure}
\epsfig{file=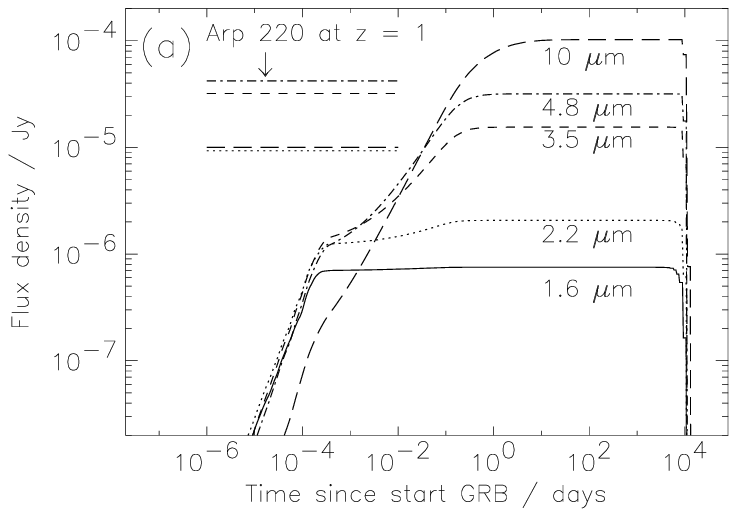}
\epsfig{file=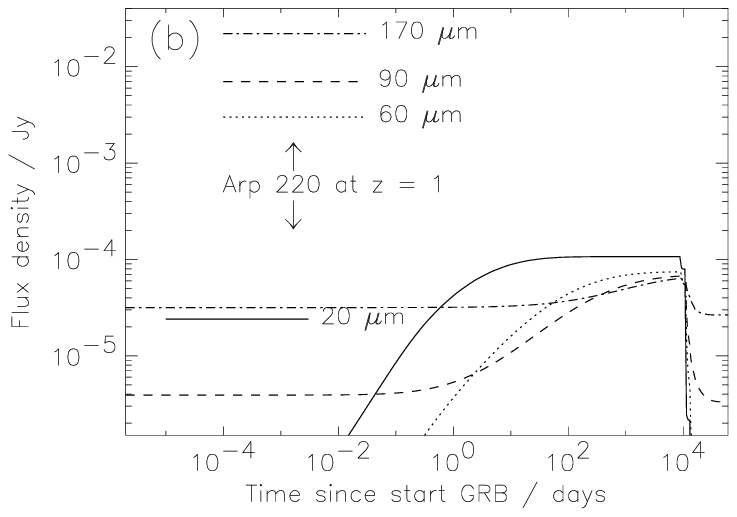}
%\vspace{-0.5cm}
\caption{\label{highlowede} (a) Emission from dust surrounding a
GRB at $z = 1$. Light curve $B_2$, which is described in Table
\ref{sims}, was assumed to heat the dust, which absorbed all the
energy of the light curve as shown in Fig.\ \ref{alllc}b. Here the
reprocessed emission at observed wavelengths of $1.6$, $2.2$, $3.5$, $4.8$
and $10\,\mu$m are plotted. (b) The corresponding emission at four
longer observed wavelengths of $20$, $60$, $90$ and
$170\,\mu$m. In both figures the flux densities at the same
wavelengths of the luminous dusty star-forming galaxy Arp 220,
redshifted to $z = 1$, are shown for comparison.}
\end{figure}

The dust emission observed from around a GRB at a given wavelength
depends on the density and profile of the surrounding dust, the peak
luminosity of the optical flash, the light curve of the GRB afterglow
and the redshift of the GRB. To illustrate the key properties of the
dust emission, a single optimistic set of model parameters was assumed.
The density profile of the dust was assumed 
to be flat with $n_{\text{H}} = 10^4$\,cm$^{-3}$. The GRB optical/UV
light curve was assumed to be $B_2$ in Table\,\ref{sims}. The redshift of
the GRB was assumed to be $z = 1$, corresponding to a luminosity
distance $d_{\text{L}} = 5.4$\,Gpc ($H_0 = 65\,$km\,s$^{-1}$\,Mpc$^{-1}$
and $q_0 = 0.5$). The GRB emission was assumed to be isotropic, not
beamed. This optimistic set of parameters also ensures an intense dust
emission signal -- more than $2400$\,M$_{\odot}$ of dust surrounds the
GRB, and nearly all the energy of the 
afterglow and optical flash is absorbed and re-emitted by the dust. In
Fig.\,\ref{highlowede} the associated dust emission predicted by the 
radiative transfer model at eight observed wavelengths between
$1.6\,\mu$m and $170\,\mu$m is plotted. A characteristic feature of
the emission from the spherical distribution of dust in this case is
that the detected signal is a top-hat function. The emission at the
high-energy end of the dust radiation spectrum is radiated predominantly by hot
grains near their sublimation temperature in the time before they
vaporize. These grains are located close to the GRB, at a distance $R
\sim 1$\,pc. Because these grains are vaporized by the intense
radiation of the optical flash, the dust emission at short wavelengths
starts to decrease about
$t\,\gtrsim\,2\,(1+z)\,R/c\,\approx\,6\,(1+z)$\,years after the GRB is
detected. The emission at longer wavelengths is predominantly due to
cooler dust, which is further away from the GRB. The dust cloud in
this model was assumed to have a maximum radius $R_{\text{max}} = 3.67$
pc, and so the emission decreases rapidly after $t > 2\,(1+z)\,
R_{\text{max}}/c \approx 22 \,(1+z)$\,years. The dust emission spectrum
is predicted to peak at a rest-frame wavelength of about $8\,\mu$m. 

\begin{figure}
\begin{center}
\epsfig{file=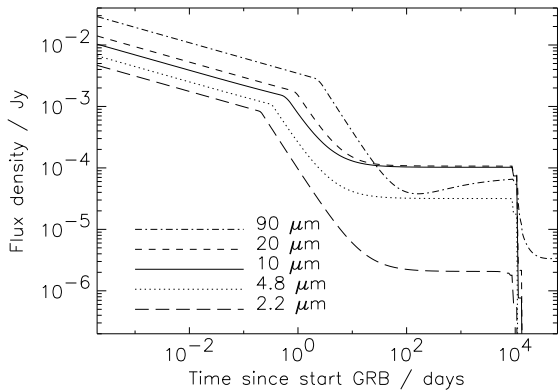}
\caption{\label{deinc} Emission from dust surrounding a GRB at $z =
1$, including emission from the GRB afterglow at five observed
wavelengths. The light curve that heated the dust is light curve
$B_2$ as shown in Fig.\,\ref{highlowede} (see Table\,\ref{sims} 
for its characteristics).}
\end{center}
\end{figure}

The dust emission predicted at the observed wavelength $170\,\mu$m at
$t < 50$\,days is due to the dust surrounding the GRB radiating at its
original temperature in the interstellar radiation field of the GRB
host galaxy, which is assumed to be $30\,$K. If the dust temperature
assumed is different, then the flux changes by only a small factor at
this wavelength. At rest-frame wavelengths longer than $100\,\mu$m, at
which the spectral energy distribution of dust at $30\,$K peaks, there
is very little excess emission due to additional heating by the GRB.

For comparison, the expected flux densities of an ULIRG similar to Arp
220 (Klaas et al.\ 1997), at a redshift $z=1$, is also plotted in
Fig.\ \ref{highlowede}. The intensity of the emission from dust heated
by the GRB at rest-frame wavelengths shorter than about $2.5 \,\mu$m
is expected to be less than that from starlight in such a
galaxy. However, the dust emission from the GRB at wavelengths of
about $5\,\mu$m can be an order of magnitude greater than the emission
from the galaxy. At rest-frame wavelengths longer than about
$25\,\mu$m, infrared emission from dust heated by star formation at
temperatures $T \simeq 40\,$K rises to dominate the
spectrum. Therefore, the dust emission from a GRB within a galaxy like
Arp 220 at $z=1$ could best be observed at mid-infrared wavelengths
between $5\,\mu$m and $20\,\mu$m. Note, however, that Arp 220 is a
very luminous galaxy. It is likely that the host galaxy of a GRB will
typically be less luminous, which would improve the observability of
the GRB dust emission. Also, the infrared emission of a galaxy will be
much more extended than emission from the surroundings of a GRB and
could be distinguished if high spatial resolution observations were
available, such as would be possible using either the Atacama Large
Millemeter Array (ALMA) or a space-borne far-infrared interferometer
like {\em SPECS} (see section\,\ref{obs} for more details).

\subsection{The infrared GRB synchrotron afterglow}

In Fig.\,\ref{highlowede} only dust emission from the environment of the
GRB was plotted. However,
it is likely that in the first few days after the GRB, the observed
spectrum at mid-infrared wavelengths will be dominated by the GRB
afterglow. In Fig.\,\ref{deinc} the dust emission at five different
wavelengths is plotted, including the emission from
the GRB afterglow. The light curves of the GRB dust emission are
the same as plotted in Fig.\,\ref{highlowede}. 
It is assumed that the infrared afterglow is not
attenuated by dust surrounding the GRB, however, at the shortest
infrared wavelengths, dust along the line of sight could
reduce the flux density of the afterglow significantly, as shown in
Table\,\ref{aandtau}. After a few days, the afterglow emission is decreasing as
$t^{-1.375}$ (Sari et al.\ 1998). All the light curves shown in
Fig.\,\ref{deinc} are dominated by the afterglow in the first few weeks
after the start of the GRB. After the first few weeks, the light curve
flattens as the afterglow emission decreases below the level of
the heated dust emission, which then dominates the light curve at late
times. If the flux density of the dust emission is ten 
times lower than that assumed in this model, then the afterglow will
dominate the spectrum for $10^{1/1.375} \approx 5$ times longer. In
other less optimistic models the GRB dust emission is typically lower
and so the afterglow will dominate the spectrum for a longer period.

\subsection{Dust emission in various environments}

\begin{figure*} 
\hspace{-0.5cm}
\epsfig{file=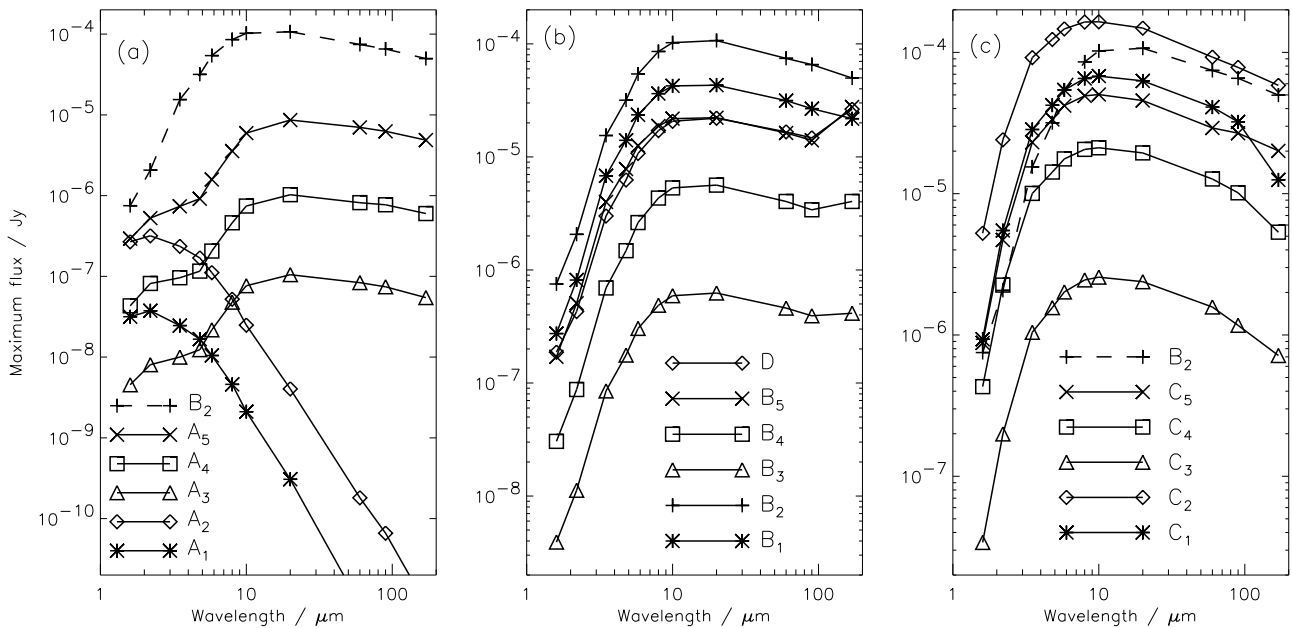}
\caption{\label{waveflux} (a)  The peak intensity of the emission spectrum
of dust heated by an optical flash with $L_{\text{peak}} =
10^{49}$\,erg\,s$^{-1}$ and an afterglow of a GRB at $z=1$. The dust
parameters are described in Table\,\ref{sims}. For comparison the spectrum
of the more powerful dust emission in
model $B_2$ is also plotted. (b) The peak intensity
of the emission spectrum of dust heated by an optical flash with
$L_{\text{peak}} = 10^{48}$\,erg\,s$^{-1}$ and an afterglow of a GRB at
$z=1$ in different dusty environments, described in Table\,\ref{sims}.
Curve $D$ has the same parameters as curve $B_5$, except that
the optical flash is not a Gaussian but a top-hat. (c) The peak
intensity of the emission spectrum of dust heated by an afterglow of a
GRB at $z=1$ with no optical flash, described by parameters 
$C_1$ to $C_5$ in Table\,\ref{sims}. For
comparison the spectrum of the dust emission in model $B_2$
is also plotted.}
\end{figure*}

The results above showed that the visibility of the optical afterglow
can depend strongly on the GRB environment. Different amounts of
energy are absorbed by different distributions of dust, thus influencing
the infrared emission. In Fig.\,\ref{waveflux}a the peak flux of the
dust emission spectrum is calculated for the $A$-series 
models listed in Table\,\ref{sims} at
$z=1$. The maximum  flux is expected to occur
roughly $(1 + z)$ years after the  GRB is detected. The time profile
of the dust emission is similar to that displayed in
Fig.\,\ref{highlowede}. The spectrum of the dust emission described
in section \ref{onesource} is plotted for comparison. In the case of
flat dust density profiles, almost all the dust is vaporized, and so the
spectrum is dominated by emission from dust near the sublimation
temperature, peaking at a wavelength about $(1+z)\,\mu$m. In the other
cases, the observed emission peaks at a longer observed wavelength
close to $10\,(1+z)\,\mu$m.

Because the optical flash is expected to destroy a large amount of
dust in the $A$-series light curve models, the intensity of the dust
emission is less than that predicted in a model with a less luminous
optical flash. This is illustrated in Fig.\,\ref{waveflux}b, in which
the dust emission spectrum is plotted for dust heated by the same GRB
afterglow, but with an optical flash that is ten times weaker,
represented by the $B$-series models, see Table\,\ref{sims}. Light
curve $D$ has the same parameters as light curve $B_5$, except that
the optical flash is this time described by a top-hat. Since in the
model the total energy radiated in the optical flash is greater, more
dust is destroyed and so the flux density of the dust emission is
smaller. Because less dust is destroyed in the $B$-series models as
compared with the $A$-series models, the intensity of the dust
emission is greater in the $B$-series models for a given density
profile.

The emission is expected to be even greater if the dust is heated by a
GRB with an afterglow but no optical flash, as shown for the
$C$-series models in Fig.\,\ref{waveflux}c. If the peak luminosity of
the optical flash is small, then almost no dust is destroyed and so
the dust emission is intense. Except for models $A_1$ and $A_2$, for
which all the dust is destroyed by the optical flash, dust with a flat
density profile emits more radiation per unit mass than dust with an
isothermal profile. To obtain the greatest dust emission from a GRB,
an optical flash with a low peak luminosity, a high-density
environment and a flat density profile are all required.

\subsection{Dust emission and the redshift of the GRB}

\begin{figure}
\epsfig{file=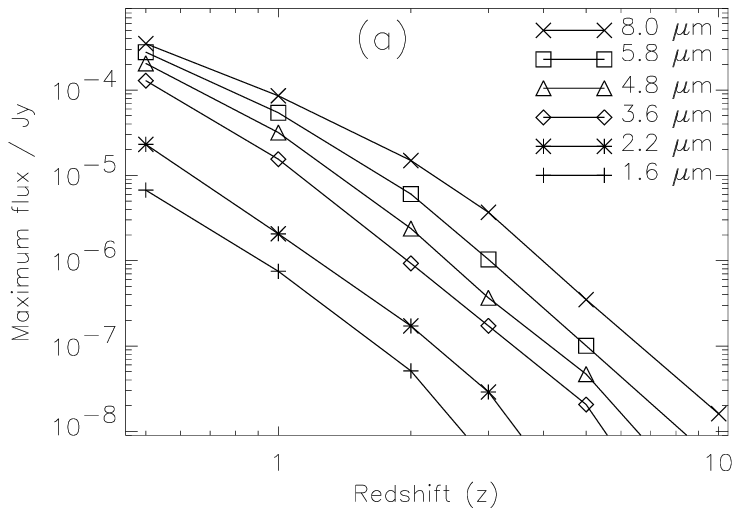}
\epsfig{file=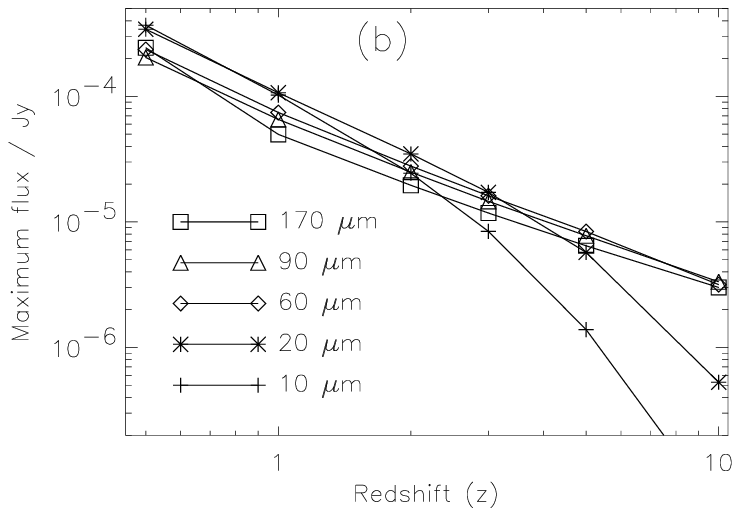}
\caption{\label{highlowez} (a) The maximum flux density of
dust emission from a GRB as a function of redshift at several
different observed wavelengths model $B_2$. The flux density
decreases rapidly as the redshift increases, between redshifts 1 and 2
the flux density decreases by a factor of about ten. The top four
curves are plotted for wavelengths observable using the {\em
SIRTF}--IRAC camera. The maximum redshifts at which the IRAC camera
could detect the dust emission with a signal to noise ratio of 5
within 1 hour of integration time are $1.7$, $2.1$, $1.8$ and $2.0$
for the $3.6$, $4.5$, $5.8$ and $8.0$ bandpass respectively. (b) The
maximum flux density of dust emission from the same GRB  at several
longer observed wavelengths. The MIPS instrument on {\em SIRTF} has
three observing bands, at $24$, $70$ and $160\,\mu$m. The maximum
redshifts at which MIPS could detect the dust emission with a
signal-to-noise ratio of 5 within 1 hour of integration time are
$0.8$, $0.4$ and $0.1$ for the $24$, $70$ and $160\,\mu$m bandpass
respectively. Confusion noise will limit the depth of observations at
the longest wavelengths, $70$ and $160\,\mu$m (Blain 1999 and
references therein).}
\end{figure}

In Fig.\,\ref{highlowez} the predicted maximum intensity of emission
by dust heated by a GRB, with a model $B_2$ optical flash afterglow,
are plotted in various bands as a function of redshift. In this
optimistic model there is strong evolution with redshift: 
the predicted flux density decreases rapidly as the redshift
increases, especially at the short wavelengths. At longer wavelengths
($\lambda \gtrsim 20\,\mu$m) the flux density decreases more slowly, as the
observed wavelength approaches the peak of rest-frame spectrum. 

\section{Observability of dust emission from GRBs}
\label{obs}

The observed wavelengths $3.5$, $4.8$, $5.8$ and $8.0\,\mu$m in
Fig.\,\ref{highlowez}a correspond to bandpasses of the InfraRed Array
Camera (IRAC) on board the {\em Space Infrared Telescope Facility}
({\em SIRTF}$\,$\footnote{Up-to-date information about SIRTF can be
found at http://sirtf.caltech.edu/}). The Multiband Imaging Photometer
for {\em SIRTF} (MIPS) has bandpasses at $24$, $70$ and
$160\,\mu$m. In 1 hour of integration time, {\em SIRTF} could detect
the near-infrared dust emission from a GRB at a redshift $z \lesssim
2$. At the longer wavelengths observable with the MIPS instrument, it
is unlikely that {\em SIRTF} will detect dust emission from a GRB at a
redshift $z \gtrsim 1$, because the emission is too weak. Also, at the
longest wavelengths observable for the MIPS instrument, $70$ and
$160\,\mu$m, confusion noise from unresolved distant galaxies will
increase the minimum detection depth, reducing the maximum redshift at
which dust emission could be observable (see Blain 1999). Because the
GRB dust emission at far-infrared and submillimetre wavelengths is
expected to be much less intense than that at near-infrared
wavelengths, facilities operating in those wavebands, like the future
Stratospheric Observatory for Infrared Astronomy (SOFIA\footnote{see
http://sofia.arc.nasa.gov/}) and the {\em Far InfraRed and
Submillimetre Telescope} ({\em FIRST--Herschel}$\,$\footnote{The
homepage of {\em FIRST} is
http://astro.estec.esa.nl/SA-general/Projects/First/first.html}), are
likely to be unsuccessful in detecting dust emission from high
redshift GRBs. However, it should be possible to resolve the GRB dust
emission using high spatial resolution facilities like
ALMA\footnote{Up-to-date information about ALMA can be found at the
web page http://www.alma.nrao.edu/} and the {\em SPECS} (see Leisawitz
et~al.\ 1999, 2000). In cases where the optical afterglow is heavily
obscured by dust and thus invisible to optical searches, ALMA could
provide a very accurate position for the GRB afterglow, once the GRB
position is known to within a few arcseconds. Accurate positions of
optically obscured afterglows are currently provided by the VLA at
radio frequencies, as shown by Frail et al.\ (2000).  Such accurate
positions can be provided by the {\em Swift} satellite\footnote{The
homepage of {\em Swift} with the latest updates can be found at
http://swift.sonoma.edu/}, which will locate a GRB precisely at X-ray
energies within 2.5$\,$arcseconds (Gehrels et~al.\ 1999). Already, a
fading counterpart to a GRB with a flux density of several mJy has
been detected at $850\,\mu$m (Smith et~al.\ 1999). In a ten-minute
observation ALMA will have a $5\sigma$  detection limit of about
0.2\,mJy at a wavelength of $870\,\mu$m. This will be deep enough to
detect the submillimetre-wave afterglow of a GRB as far away as a
redshift of 10 within a day of the start of the GRB, while the
afterglow of a GRB at a redshift of 5 would be detectable for three
weeks.

\section{Discussion}
\label{discussion}

In this paper the visibility of optical/UV GRB emission absorbed by
dust and the intensities of the reprocessed radiation emitted by dust
were predicted. The heating effect of GRB afterglow and prompt optical
flash was calculated in a model whose parameters are consistent with
the relatively sparse existing observations. A radiative
transfer model was used to simulate the interaction between GRB
radiation and dust grains; the temperature, opacity and
sublimation rate of the dust grains in the environment of a GRB, and
the infrared emission from this dust was calculated. Note that in the
model the GRB emission was assumed to be isotropic and that dust
destruction by X-rays was neglected. If GRBs are beamed and/or grains
are predominantly destroyed by X-rays, then the dust emission as
computed in the model could be reduced significantly.  

The simulations showed that the visibility of the optical flash and
afterglow of a GRB depends strongly on the characteristics of both the
surrounding dust and the GRB emission itself. If a GRB is surrounded
by a large amount of dust ($M_{\text{dust}} \gtrsim
100$\,M$_{\odot}$), then it is unlikely that an optical afterglow will
be observed, unless the GRB is accompanied by a very powerful optical
flash with a peak luminosity $L_{\text{peak}} \gtrsim
10^{49}$\,erg\,s$^{-1}$, which would destroy almost all the dust along
the line of sight within a few seconds of the GRB. Because of the
large-scale destruction of dust grains in this scenario, the intensity
of the associated dust emission would be quite weak. If the optical
flash associated with a GRB is less powerful, then less dust is
destroyed. The opacity is thus higher, and so it is possible that the
energy in the afterglow is almost completely absorbed, and that no
late-time optical emission from the GRB would be observed. However,
the reradiated energy might be detectable.  In particular, the IRAC
camera on {\em SIRTF} could detect the near-infrared dust emission in
such cases out to redshifts of about $2$, if the GRB could be located
to within its $5-$arcmin field of view. Because the light curve of
this dust emission does not have any specific temporal structures, and
could therefore be confused with the flux density of dust radiation
from the general interstellar medium of the host galaxy, an accurate
position and flux density of the GRB afterglow would need to be provided by
other observations. At present these can be carried out with the
VLA. In the future high-resolution submillimetre-wave observations
with ALMA will be able to confirm that the origin of the infrared
energy was a GRB.

A combination of observations is needed to test whether a GRB is
surrounded by a large amount of dust. First of all, the lack of a
detection of an afterglow in the optical waveband could be a first
indication that there is a large mass of dust near the GRB. If
radio observations provide an accurate position for the afterglow,
then the {\em SIRTF} mid-infrared telescope could be used to try to
detect the reprocessed dust emission directly. However, only in the
most  optimistic scenarios, in which more than $10^3$\,M$_{\odot}$ of
dust is present and the optical flash of the GRB is weak, is infrared
emission likely to be observed from dust heated by the optical flash
and afterglow of the GRB. If either the GRB emission is beamed, or
X-ray grain destruction is important, or less than $10^2$\,M$_{\odot}$
of dust surrounds the GRB, then the infrared dust emission is likely
to fall below the detection limits of {\em SIRTF} unless the GRB is at
a low redshift $z < 0.5$.

The results of the model can be compared with the results of existing
observations for optical flashes from GRBs. A powerful optical flash
was observed to accompany GRB990123, detected twenty seconds after the
start of the GRB (Akerlof et al.\ 1999). From the observed redshift
$z=1.6$ (Galama et~al.\ 1999b), the peak luminosity was calculated to
be $\sim 10^{50}$\,erg\,s$^{-1}$. This flash could easily have
vaporized more than $10^3$\,M$_{\odot}$ of dust around the GRB within a few
seconds. The late-time afterglow for this GRB is thus unlikely to have
been attenuated by dust surrounding the GRB, and so the dust emission
in this case was probably very low. Attempts to detect the prompt
optical flash from other GRBs have provided only upper limits (Akerlof et al.\
2000). This does not necessarily  imply that the optical
emission in those cases was extinguished by surrounding dust. However, if the
peak luminosity of the optical flash of a GRB is well correlated with
either gamma-ray fluence or gamma-ray peak flux, then in two of the
observed cases the optical flash would have $L_{\text{peak}} \simeq
10^{49}$\,erg\,s$^{-1}$. This luminosity is great enough to sublime
most of the surrounding dust, and thus lead to the detection of optical
emission. Even if these GRBs occurred in very dusty environments, then
the lack of optical emission indicates that the peak luminosity of the
optical flash is probably not strongly correlated with either gamma-ray fluence
or gamma-ray peak flux.

\section*{Acknowledgments} 
We would like to thank the referee R.\,A.\,M.\,J.\ Wijers, for his
helpful and useful comments, and for pointing out the X-ray dust
destruction results of Fruchter et~al.\ (2001).  
BPV acknowledges the support of a scholarship from the Netherlands
organization for international cooperation in higher education, to
study in Cambridge (Venemans 2000).  
AWB, Raymond and Beverly Sackler Foundation Research Fellow, thanks
the Foundation for generous financial support as part of their
Deep Sky Initiative Programme at the IoA.


\begin{thebibliography}{99}
\bibitem{} Aitken D.\,K., Moore T.\,J.\,T., Roche P.\,F., Smith
C.\,H., Wright~C.\,M., 1993, MNRAS, 265, L41
\bibitem{} Akerlof C.\ et~al., 1999, Nat, 398, 400
\bibitem{} Akerlof C.\ et~al., 2000, ApJ, 532, L25
\bibitem{} Blain A.\,W., 1999, in Weymann R., Storrie-Lombardi L.,
Sawicki M., Brunner R., eds, Photometric Redshifts and the Detection of
High Redshift Galaxies, ASP Conf.\ Ser.\ Vol. 191, 255, ASP, San Francisco
\bibitem{} Bloom J.\,S. et~al., 1999a, Nat, 401, 453
\bibitem{} Bloom J.\,S. et~al., 1999b, ApJ, 518, L1
\bibitem{} Bloom J.\,S., Kulkarni S.\,R., Djorgovski S.\,G., 2001, AJ,
submitted (astro-ph/0010176)
\bibitem{} Cardelli J.\,A., Clayton G.\,C., Mathis J.\,S., 1989,
ApJ, 345, 245
\bibitem{} Costa E.\ et~al., 1997, IAU Circ., 6572, 1
\bibitem{} D{\'e}sert F.\,X., Boulanger F., Puget J.\,L., 1990,
A\&A, 237, 215
\bibitem{} Draine B.\,T., 2000, ApJ, 532, 273
\bibitem{} Draine B.\,T., 2001,
http://www.astro.princeton.edu/ $\sim$draine/dust/dust.diel.html 
\bibitem{} Draine B.\,T., Lee H.\,M., 1984, ApJ, 285, 89
\bibitem{} Esin A.\,A., Blandford R., 2000, ApJ, 534, L151
\bibitem{} Evans A., 1993, The dusty universe. Wiley, New York
\bibitem{} Frail D.\,A.\ et~al., 2000, in Kippen R.\,M., Mallozzi R.\,S.,
Connaughton V., eds., AIP Conf. Proc. 526, Gamma-Ray Bursts: Fifth
Huntsville Conference. American Institute of Physics, New York
\bibitem{} Frail D.\,A. et~al., 2001, Nature, submitted (astro-ph/0102282)
\bibitem{} Fruchter A.\,S. et~al., 1999, ApJ, 516, 683
\bibitem{} Fruchter A.\,S., Krolik J.\,H., Rhoads J.\,E., 2001, preprint
\bibitem{} Galama T.\,J., Wijers R.\,A.\,M.\,J., 2001, ApJ, 549, L209
\bibitem{} Galama T.\,J.\ et~al., 1999a, A\&AS, 138, 451
\bibitem{} Galama T.\,J.\ et~al., 1999b, Nat, 398, 394
\bibitem{} Gehrels N., Swift Science Team, 1999, BAAS, 195, 9208
\bibitem{} Greiner J., 2001, http://www.aip.de/$\sim$jcg/grbgen.html
\bibitem{} Groot P.\,J.\ et~al., 1997, IAU Circ., 6584, 1
\bibitem{} Guhathakurta P., Draine B.\,T., 1989, ApJ, 345, 230
\bibitem{} Hanlon L.\ et~al., 2000, A\&A, 359, 941
\bibitem{} Hildebrand R.\,H., 1983, QJRAS, 24, 267
\bibitem{} Hobson M.\,P., Padman R., 1993, MNRAS, 264, 161
\bibitem{} Hobson M.\,P., Scheuer P.\,A.\,G., 1993, MNRAS, 264, 145
\bibitem{} Huang Y.\,F., Dai Z.\,G., Lu T., 2000, A\&A, 355, L43
\bibitem{} Katz J.\,I., 1994, ApJ, 432, L107
\bibitem{} Klaas U., Haas M., Heinrichsen I., Schulz B., 1997, A\&A,
325, L21
\bibitem{} Klebesadel R.\,W., Strong I.\,B., Olson R.\,A., 1973, ApJ,
182, L85 
\bibitem{} Leisawitz D., Mather J.\,C., Moseley S.\,H., Xiaolei Z.,
1999, Ap\&SS, 269, 563
\bibitem{} Leisawitz D., Mather J.\,C., Blain A.\,W., Langer W.\,D.,
Moseley S.\,H., Yorke H.\,W., 2000, BAAS, 197, 1409
\bibitem{} Lamb D.\,Q., Castander F.\,J., Reichart D.\,E., 1999,
A\&AS, 138, 479
\bibitem{} MacFadyen A.\,I., 2001, in Kippen R.\,M., Mallozzi R.\,S.,
Connaughton V., eds., AIP Conf. Proc. 526, Gamma-Ray Bursts: Fifth
Huntsville Conference. American Institure of Physics, New York
\bibitem{} MacFadyen A.\,I., Woosley S.\,E., 1999, ApJ, 524, 262
\bibitem{} Mathis J.\,S., 1990, ARA\&A, 28, 37
\bibitem{} Mathis J.\,S., Rumpl W., Nordsieck K.\,H., 1977, ApJ,
217, 425
\bibitem{} M{\'e}sz{\'a}ros P., Gruzinov A., 2000, ApJ, 543, L35
\bibitem{} M{\'e}sz{\'a}ros P., Rees M.\,J., 1993, ApJ, 418, L59
\bibitem{} M{\'e}sz{\'a}ros P., Rees M.\,J., 1997, ApJ, 476, 232
\bibitem{} Mezger P.\,G., Mathis J.\,S., Panagia N., 1982, A\&A,
105, 372
\bibitem{} Paczynski B., 1995, PASP, 107, 1167
\bibitem{} Perna R., Aguirre A., 2000, ApJ, 543, 56
\bibitem{} Piran T., 1999, PhR, 314, 575
\bibitem{} Reichart D.\,E., 1999, ApJ, 521, L111
\bibitem{} Sahu K.\,C.\ et~al., 1997, Nat, 387, 476
\bibitem{} Sari R., Piran T., 1999, A\&AS, 138, 537
\bibitem{} Sari R., Piran T., Narayan R., 1998, ApJ, 497, L17
\bibitem{} Siebenmorgen R., Kr{\"u}gel E., Mathis J.\,S., 1992,
A\&A, 266, 501 
\bibitem{} Smith I.\,A.\ et~al., 1999, A\&A, 347, 92
\bibitem{} Sokolov V.\,V. et al., 2001, A\&A, 372, 438
\bibitem{} Venemans B.\,P., 2000, MPhil thesis, Univ. Cambridge
\bibitem{} Waxman E., Draine B.\,T., 2000, ApJ, 537, 796
\bibitem{} Witt A.\,N., Oliveri M.\,V., Schild R.\,E., 1990, AJ, 99,
888
\end{thebibliography}
\end{document}